\documentclass[11pt, twoside]{article}
\usepackage{a4wide}
\usepackage[fleqn]{amsmath}
\usepackage{amssymb, exscale}
\usepackage{epsfig}
\usepackage{soul,titlesec}
\usepackage{cite}
\usepackage[marginal]{footmisc} 
\usepackage[overload]{textcase} 
\usepackage{subfig}

\makeatletter
\renewcommand\thesection{\oldstylenums{\arabic{section}}}
\renewcommand\thesubsection{\thesection.\oldstylenums{\arabic{subsection}}}
\renewcommand\thesubsubsection{\thesubsection.\oldstylenums{\arabic{subsubsection}}}

\@addtoreset{equation}{section}

\newcommand{\sectionname}{Section}
\newcommand{\marktotoc}[1]{\renewcommand{\sectionname}{#1}}
\renewenvironment{appendix}
{\renewcommand{\sectionname}{Appendix}%
  \addtocontents{toc}{\protect\marktotoc{Appendix}}%
  \setcounter{section}{0}%
  \renewcommand\thesection{{\sc\alph{section}}}%
}
{\addtocontents{toc}{\protect\marktotoc{Section}}}

\soulregister{\MakeTextUppercase}{1}
\soulregister{\MakeTextLowercase}{1}
\capsdef{/cmr/m/n/11-13}{\scshape}{0.13em}{0.5em}{0.33em}
\capsdef{/cmr/m/n/9-11}{\scshape}{0.05em}{0.33em}{0.33em}

\newcommand\secformat[1]{\so{\MakeTextUppercase{#1}}}
\newcommand\subsecformat[1]{\caps{\MakeTextLowercase{#1}}}
\newcommand\subsubsecformat[1]{{\it #1}}
\titleformat{\section}[block]{}{\thesection\quad}{1em}{\secformat}[]
\titleformat{\subsection}[block]{}{\thesubsection\quad}{1em}{\subsecformat}[]
\titleformat{\subsubsection}[block]{}{\thesubsubsection\quad}{1em}{\subsubsecformat}[]

\titlespacing*{\section}{0pt}{2\baselineskip}{2\baselineskip}
\titlespacing*{\subsection}{0pt}{\baselineskip}{\baselineskip}
\titlespacing*{\subsubsection}{0pt}{\baselineskip}{\baselineskip}

\makeatother
\begin{document}
\newcommand{\osn}[1]{\oldstylenums{#1}}
\newcommand{\diag}{\mathrm{diag}}
\newcommand{\chpt}{$\chi$\sc pt\rm}
\newcommand{\lecs}{\sc lec\rm s}
\renewcommand{\Re}{\mathrm{Re}\;}
\renewcommand{\Im}{\mathrm{Im}\;}
\newcommand{\relphantom}[1]{\mathrel{\phantom{#1}}}
\newcommand{\hhv}[2]{\smash{\overset{\leftrightarrow}{v}}^{#1}_{#2}}
\newcommand{\qcd}{\sc qcd\rm}
\newcommand{\SU}{\mathrm{SU}}
\renewcommand{\L}{\mathcal{L}}
\newcommand{\T}{\mathrm{T}}
\renewcommand{\l}{\ell}
\newcommand{\Kr}[1]{K^{\mathrm{r}}_{#1}}
\newcommand{\kr}[1]{k^{\mathrm{r}}_{#1}}
\newcommand{\Lr}[1]{L^{\mathrm{r}}_{#1}}
\newcommand{\lr}[1]{l^{\mathrm{r}}_{#1}}
\newcommand{\Cr}[1]{C^{\mathrm{r}}_{#1}}
\newcommand{\clr}[1]{c^{\mathrm{r}}_{#1}}
\newcommand{\Hr}[1]{H^{\mathrm{r}}_{#1}}
\newcommand{\hr}[1]{h^{\mathrm{r}}_{#1}}
\newcommand{\bj}{\mathbf{j}}
\newcommand{\bp}{\mathbf{p}}
\newcommand{\bra}{\langle}
\newcommand{\ket}{\rangle}
\newcommand{\nn}{\nonumber}
\renewcommand{\d}{\partial}
\newcommand{\ord}[1]{O\!\left(#1\right)}
\newcommand{\comm}[2]{[#1,#2]}
\newcommand{\anticomm}[2]{\{#1,#2\}}
\newcommand{\sprod}[2]{(#1,#2)}
\newcommand{\Tr}{\operatorname{Tr}}
\newcommand{\Cl}{\operatorname{Cl_2}}
\newcommand{\tr}{\operatorname{tr}}
\newcommand{\co}{\,,}
\newcommand{\mc}{\,;}
\newcommand{\fs}{\,.}
\newcommand{\dmeas}[2]{\mathrm{d}^{#2}#1}
\def\slashi#1{\rlap{\sl/}#1}
%
\def\slashii#1{\setbox0=\hbox{$#1$}             
   \dimen0=\wd0                                 
   \setbox1=\hbox{\sl/} \dimen1=\wd1            
   \ifdim\dimen0>\dimen1                        
      \rlap{\hbox to \dimen0{\hfil\sl/\hfil}}   
      #1                                        
   \else                                        
      \rlap{\hbox to \dimen1{\hfil$#1$\hfil}}   
      \hbox{\sl/}                               
   \fi}                                         %
%
\def\slashiii#1{\setbox0=\hbox{$#1$}#1\hskip-\wd0\hbox to\wd0{\hss\sl/\/\hss}}
%
\def\slashiv#1{#1\llap{\sl/}}
\begin{center}
{\Large Integrating out the heaviest quark in $N$--flavour \chpt{}}
\vspace{4ex}

Mikhail A. Ivanov$^1$ and Martin Schmid$^2$
\vspace{2ex}

{\it $^1$Bogoliubov  Laboratory of Theoretical Physics, 
Joint Institute for Nuclear Research,\\
141980 Dubna, Russia
\vspace{1ex}

$^2$ Basler Versicherung {\small AG}, Aeschengraben 21,\\
4002 Basel, Switzerland
}
\end{center}
\caps{Abstract}
\vspace{1ex}

{\small We extend a known method to integrate out the strange quark in
  three flavour chiral perturbation theory to 
  the context of an arbitrary number of flavours. As an application, we present the
  explicit formul\ae{} to one--loop accuracy for the heavy quark mass
  dependency of the low energy constants after decreasing the number of
  flavours by one while integrating out the heaviest quark in $N$--flavour
  chiral perturbation theory.
}
\vspace{2ex}

\vspace{2ex}

\noindent\caps{pacs}: 11.30.Rd, 12.39.Fe, 13.40.Dk, 13.40.Ks

\section{Introduction}

Chiral perturbation theory (\chpt) \cite{Weinberg78,glann,glnpb} displays and exploits
transparently the symmetries of low energy \qcd. However, being an effective
field theory, it also features a myriad of low--energy constants (\lecs) that are not
fixed by symmetry, but rather have to be determined from experiment. To aid
this determination with additional constraints and gain some insight into the
heavy quark mass dependence of these \lecs, a series of
publications\cite{Moussallam:2000zf,Kaiser:2006uv,Gasser:2007sg,Gasser:2009hr,Gasser:2001un,Jallouli:1997ux,Haefeli:2007ey,Kampf:2009tk}
has appeared in the recent past that presents relations among the \lecs{}
of different versions of \chpt. This work will
contribute to this line of publications, albeit in an unusual
form, as it addresses a matching between the \lecs{} of \chpt{} with $N$ light
quarks (\chpt$_N$) and
\chpt$_{N-1}$, while the publications cited
above concentrate on three-- versus two--flavour physics. The reason for this
generalisation lies in a possible interest of the lattice community in a matching
between four-- and three-- flavour \lecs. To not repeat
ourselves with only different numbers, we chose to generalise the scope, as we
do not know if there might arise some interest in other flavour combinations in the
future.

The aim of this paper is to provide the dependence on the $N^\text{th}$ quark
mass of the \lecs{} of chiral perturbation theory of $N-1$
flavours to one--loop accuracy. This aim is achieved by the methods laid out in
\cite{tinu}, namely calculating the generating functional for \chpt${}_N$
in a limit where it describes only the physics of \chpt${}_{N-1}$. Comparing the coefficients of its local contributions with
the action of \chpt${}_{N-1}$ yields the desired matching. This method is
equivalent to evaluate and compare the Green's functions of external fields of
both theories, however, without the need for a cumbersome calculation and
comparison of multiple matrix elements in both theories. The method can be
used for higher loop calculations with only marginal complications (but at a
significantly higher computational effort).

The publication is structured as follows: after this brief introduction, the
formalism of \chpt${}_N$ is laid out in
Section~\ref{Sec:chpt} and in more detail in Appendix~\ref{app:notation}. It follows a description of the technique used in
Section~\ref{Sec:limit}. The technical details of the tree--level
(Section~\ref{app:tree}) and the loop calculations (Section~\ref{app:loop})
are devoted their own space, summarised in Section~\ref{sec:calculation}. At the end, we add Section~\ref{Sec:results}
with the results of the calculation, examples of application in
Section~\ref{sec:application} and a short summary in
Section~\ref{Sec:summary}. In Appendix~\ref{app:validity}, a reduction to the case
$N=3$ is presented as a check of the calculations.

\section{Preliminaries}
\label{Sec:chpt}

In this section, we will shortly discuss the setup of \chpt${}_N$.

Chiral perturbation theory yields a consistent and systematic
framework to explore the effects of symmetries in low energy \qcd. The
starting point is the massless \qcd--Lagrangian $\L_{\qcd{}}^0$, enriched
with couplings to external (axial--) vector fields ($a_\mu$) $v_\mu$ and (pseudo--)
scalar sources ($p$) $s$.

The leading order (Euclidean) Lagrangian reads
\begin{equation}
  \label{eq:lagr2}
  \L_2^N = \frac{F_N^2}{4}\bra u^N\cdot u^N-\chi^N_+\ket\co
\end{equation}
where the above mentioned (axial--) vector fields are part of the building
block $u_\mu^N$ and the (pseudo--) scalar fields are hidden in
$\chi_+^N$. Consult Appendix~\ref{app:notation} for the details of the
notation. Visible is one of the two leading order \lecs{} ($F_N$), while the
other ($B_N$) also hides in $\chi_+^N$.

This Lagrangian implies the equations of motion
\begin{equation}
  \label{eq:eom}
  \nabla^N_\mu u^N_\mu +\tfrac{i}{2}\tilde{\chi}^N_-=0\co
\end{equation}
with $\tilde{\chi}^N_-$ the traceless part of $\chi^N_-$ and $\nabla^N_\mu$ the
covariant derivative $\nabla^N_\mu\cdot{}=\d_\mu\cdot{}+\,[\,\Gamma_\mu^N\co\,\cdot{}\;]$.

The general form of the next--to--leading order Lagrangian $\L_4^N$ for a
generic number of flavours consists of thirteen terms $X_k^N$ and as many \lecs{}
$L_k^N$: $\L_4^N = \sum_{k=0}^{12}L_k^NX_k^N$. Here, we disregard terms that
vanish at the solution of the equation of motion \cite{renorm}, as these are
irrelevant at one loop. Note that this generic form already
shows up at $N = 4$, where in addition to the familiar structure for
$N=3$ the term proportional to $L_0^N$ is needed \cite{glnpb}, see also
(\ref{eq:L4ops}). If $N$ is smaller than four, Cayley--Hamilton
relations between the structures $X_k^N$ reduce the number of needed elements
to \osn{12} for $N = 3$ and \osn{10} for $N = 2$.

The generating functional $Z_N$ can be written in a series where the elements
are ordered by the number of loops involved in their determination. This
series is equivalent to reintroduce the old--fashioned $\hbar$ and expanding
$Z$ in powers of $\hbar$,
\[Z^N=\bar{S}_2^N+\hbar\left(\bar{S}_4^N+\tfrac{1}{2}\ln\det D^N/D_0^N\right) + \ord{\hbar^2}\co\]
where $\bar{S}_n^N$ denotes the classical action belonging to $\L_n^N$ and the
differential operator $D^N$ is obtained from the second order variation of
$S_2^N$. For details, consult (\ref{eq:diffop}).


\section{$N-1$--flavour limit}
\label{Sec:limit}

To obtain the $N^\text{th}$ quark mass dependence of the $N-1$--flavour low--energy constants,
we will use a field theoretic approach. Namely we will determine the local
contributions to the $N$--flavour generating functional in a limit of external
momenta and fields where the $N$--flavour chiral perturbation theory reduces
to the one of $N-1$ flavours. This reduction can be obtained by the three
following steps:
\begin{itemize}
\item[-] reduce the external sources to the ones of \chpt$_{N-1}$:
  $x^N=\diag(x^{N-1},0)$ for $x\in\{v_\mu;a_\mu;p\}$ and
  $s^N=\diag(s^{N-1},m)$, i.e. $m_N =m$.

  This reduction leads to a separation of the fields analogous as in the
  $\SU(3)$: there are fields that are fully described within \chpt$_{N-1}$,
  denoted by $\pi$, a field that mixes with the diagonal component of $\pi$,
  denoted by $\eta$, and the remaining \chpt$_N$--fields, denoted collectively
  by $K$.
\item[-] require that the quark masses $m_1,\ldots,m_{N-1}\ll m$. Technically, it
  is easier to put all the light quark masses even to zero, as this avoids
  complications with additional scales on which the result does not depend
  ($N$--flavour \lecs{} do no depend by definition on the $N$ lightest quark
  masses). Therefore, we will apply this technical simplification for our
  calculations. 

  As a consequence, the tree--level mass--squares of the particles
  simplify  drastically:
  \begin{equation}
    \label{eq:tree_mass}
    \bar{M}_\pi^2 = 0\co\qquad\bar{M}_K^2 = B_Nm\co\qquad\bar{M}_\eta^2 = 2\tfrac{N-1}{N}B_Nm\fs
  \end{equation}
  
\item[-] only consider processes with a low invariant $q^2$, such that virtual
  $\eta$-- or $K$--particles cannot go on--shell: $q^2\ll B_Nm$. As a consequence,
  the heavy particle loop content is analytic and can be expanded in a series
  of $q^2/B_Nm$ around $q^2 = 0$, leading to local contributions in the generating functional.
\end{itemize}

The next step in the matching process is to define appropriate counting
criteria. Apart from the counting in powers of external momentum, determining
the operators of order $q^{2n}$ belonging to $\L^N_{2n}$\footnote{Note that we
use the standard counting of \cite{glann}. Other rules have been discussed,
consult
e.g. \cite{DescotesGenon:1999uh,DescotesGenon:2002yv,DescotesGenon:2003cg}. The
literature can be tracked from \cite{DescotesGenon:2007bs}. However, these differences do not matter here, as we are purely
interested in an algebraic relation between operators of different flavour
number, irrespective of their momentum counting.}, we will consider the \lecs{}
belonging to $\L_{2n}$ to be of order $\hbar^{n-1}$, as the pertinent tree--level
contribution to the generating functional is of the same order. This counting
is only consistent if we further assume the quantity $B_N m$ to be
of order $\hbar^{-1}$, as the \lecs{} will be written as an expansion in the
quark mass $m$. Since every further term in this expansion is obtained by a
higher loop calculation, the coefficient will be of a higher order in
$\hbar$. For all these terms in the series to be of the same order in
$\hbar$, the quantity $B_Nm$ must hence be of order $\hbar^{-1}$. This series
representation will reveal the quark mass dependency of the
\chpt$_{N-1}$--\lecs. We will work out the
relations up to order $\hbar$.

Once these initial questions are settled, one first has to translate the
operators appearing in the $N$--flavour theory into the ones of
$N-1$--flavours. This is done by solving the equations of motion of the
particles not present in the $N-1$--flavour variant. Details on this process
are given in the next section and in Appendix~\ref{app:tree}.

Then, one has to extract the (now, due to the limiting process) local
contributions to the generating functional of the loop diagrams. The details
to this calculation are given below and in Appendix~\ref{app:loop}.  Once all
these contributions are known, the proper matching process can be accomplished
by comparing the coefficients of a given $N-1$--flavour operator.


\section{Calculation}
\label{sec:calculation}

In this section, the necessary steps of the calculation are sketched. A
detailed description can be found in Appendices~\ref{app:tree}
and~\ref{app:loop}.  

\subsection{Tree-level}
The tree level calculation boils down to solve the equation of motion
\cite{Brown}. We will therefore express the solutions of the $N$--flavour
fields (within the limits set out in the preceding section) in the language of
the the building blocks of \chpt${}_{N-1}$. Hence, a translation table from
the building blocks of \chpt${}_N$ to those of \chpt${}_{N-1}$ is generated. 

The first observation to make is that, in the $N-1$--flavour limit, the
solution of the equations of motion for the $K$--fields is trivial. The
argument runs along current conservation and leads to the solution
\begin{equation}
  \label{eq:tree_form}
  {u^N}=u^\pi u^\eta\fs
\end{equation}
Hence the solution to the equation of motions is split into two commuting
parts depending solely on $\pi$--fields (the part $u^\pi$) and on
$\eta$--fields (the part $u^\eta$). As $u^\pi$
is an element of $\SU(N-1)$, this solution immediately leads to a
representation of the $N$--flavour building blocks in terms of the building
blocks of \chpt${}_{N-1}$ and the $\eta$--field. It can further be shown that
at the
one--loop level of the perturbation theory, the $\pi$--fields coincide with the
fields of \chpt${}_{N-1}$, hence the translation for the $K$-- (trivially) and
the $\pi$--fields is already complete. 

To find an expression for the $\eta$--field in terms of \chpt${}_{N-1}$ and its
sources, it suffices to re--express $\L_2^N$ in the representation as found
above and extract the equation of motion for $\eta$, which can be readily
solved. One obtains the solution
\begin{equation}
  \label{eq:eta}
  \eta=-i\frac{F_N}{8B_{N-1}m}\sqrt{\frac{2N}{(N-1)^3}}\bra\chi^\pi_-\ket+O(q^4)\fs
\end{equation}

The representations of the building blocks are
\begin{equation}
  \label{eq:build1}
  \begin{split}
    u^N_\mu&=u^\pi_\mu-\tfrac{1}{F_N}\lambda_\eta\d_\mu \eta\co\\
    \chi^N_\pm&=\frac{B_N}{B_{N-1}}\left(\chi^\pi_\pm\cos\alpha-i\chi_\mp^\pi\sin\alpha\right)+4B_Nm\,e_{NN}
    \begin{cases}
       \phantom{i}\cos (N-1)\alpha & \quad{\chi}_+^N\\
      i\sin (N-1)\alpha & \quad{\chi}_-^N
    \end{cases}\co\\
  \end{split}
\end{equation}
with $\alpha=\sqrt{2/[N(N-1)]}\,\eta/F_N$ and operators $X^\pi$ denote $X$ evaluated
with the fields $u^\pi$ and in the external fields only the $\SU(N-1)$--part
being different from zero, $B_N$ replaced by $B_{N-1}$. The only nonzero entry
of the matrix' $e_{NN}$ is a 1 in the lower right corner and $\lambda_\eta
=\sqrt{\frac{2}{N(N-1)}}\;\diag\big(1_{N-1},1-N\big)$. As can be seen from
(\ref{eq:eta}), $\alpha$ is a quantity of order $q^2$, hence the trigonometric
functions can be expanded up to the required order to obtain an explicit
expression.

\subsection{Loops}
For the loop contributions, it suffices to determine the terms becoming local
when applying the $N-1$--flavour limit to 
\[Z_{1\,\mathrm{loop}}^N=\tfrac{1}{2}\ln\frac{\det D^N}{\det D^N_0}\fs\]

This determinant can be splitted into massive and massless contributions in
the following way \cite{Nyffeler:1994ph} (the index of $D^N$ denoting the
subspace to consider):
\begin{equation}
  \label{eq:det}
  \ln\det D^N = \ln\det D_{\pi}+\ln\det D_\eta+\ln\det
  D_K+\ln\det(1-D_\pi^{-1}D_{\pi\eta}D_\eta^{-1} D_{\eta\pi})\co
\end{equation}
The first term containing only $\pi$--fields can be neglected, as it will produce
exclusively non--local contributions to the generating functional. The next two determinants describe tadpoles with insertions where only
particles of identical masses run in the loop: either $K$-- or
$\eta$--particles. Diagrams of this type are most efficiently calculated using the
heat--kernel formalism, details are given in Appendix~\ref{app:loop}. The last term
describes the loop mixing contributions between the $\pi$-- and
$\eta$--fields. For obtaining local contributions, only one massless
$\pi$--propagator can appear in the diagram. The massive $\eta$--propagators can
again be expanded via the heat--kernel formalism, but at this level of the
counting we only need the leading free propagator.

All in all, the local contribution to $Z_\text{1loop}^N$ is of the form 
\begin{equation}
  \label{eq:1loop:1}
  \tfrac{1}{2}\ln\frac{\det D^N}{\det D_0^N}^\text{loc}=\int
  \mathrm{d}^dx\,(\L_\eta^N+\L_K^N+\L_{\eta\pi}^N)
\end{equation}
with
\begin{equation}
  \label{eq:1loop:2}
  \begin{split}
    \L_\eta^N &=\tfrac{1}{4N(N-1)}F_1(\bar{M}_\eta^2)\left[\tfrac{B_N}{B_{N-1}}\bra{\chi}_+^l\ket-\tfrac{1}{8B_Nm}\left(1-\tfrac{2}{(N-1)^2}\right)X_7^{N-1}\right]\\
&\relphantom{=}+\tfrac{1}{16N^2(N-1)^2}F_2(\bar{M}_\eta^2)X_6^{N-1}\co\\
    \L_K^N &= -\tfrac{1}{4}F_1(\bar{M}_K^2)\left(\bra{u}_\mu^l{u}_\mu^l\ket-\tfrac{B_N}{B_{N-1}}\bra{\chi}^l_+\ket+\tfrac{N-3}{8B_Nm(N-1)^2}X_7^{N-1}\right)\\
&\relphantom{=}+\tfrac{1}{48}F_2(\bar{M}_K^2)\Big(\tfrac{1}{2}X_0^{N-1}+X_3^{N-1}+3X_5^{N-1}+\tfrac{3}{2}X_8^{N-1}\\
&\relphantom{=+\tfrac{1}{24}F_2(\bar{M}_K^2}+2X_9^{N-1}-2X_{10}^{N-1}-X_{11}^{N-1}+3
X_{12}^{N-1}\Big)\co\\
    \L_{\eta\pi}^N &= \tfrac{-1}{8N(N-1)}F_2^1(\bar{M}^2_\eta)
\left(\tfrac{1}{N-1}X^{N-1}_6-X^{N-1}_8-2X^{N-1}_{12}\right)\mc
  \end{split}
\end{equation}
with loop integrals denoted by $F_n^m(z) = (2\pi)^{-d}\int\!\mathrm{d}\ell\,\ell^{-2m}(z+\ell)^{m-n}$. These can be treated via the
standard $\overline{\text{MS}}$--scheme, customary in \chpt{} and also
described in Appendix~\ref{app:loop}.

\section{Results}
\label{Sec:results}

We compare terms with the operator $\bra u^\pi\cdot u^\pi\ket$ in
both theories to extract a matching for $F_{N-1}$. Once this is done, we
compare terms with the operator $\bra\chi_+^\pi\ket$. On the $\SU(N)$--side,
they are all accompanied by the factor $B_N/B_{N-1}$. Bringing the denominator
to the other side and inserting the result for $F_{N-1}$, one obtains the
matching for $B_N$. We hence get for the \lecs{} of $\L_2^{N-1}$ to
next--to--leading order
\begin{equation}
  \label{eq:FB}
  \begin{split}
    F_{N-1} &= F_N\left(1-\frac{\mu_K}{F_N^2} + 8\frac{B_Nm}{F_N^2}L_4^\mathrm{r}{}^N\right)\co\\
    B_{N-1} &=
    B_N\left[1-\tfrac{2}{N(N-1)}\frac{\mu_\eta}{F_N^2}-\frac{16B_Nm}{F_N^2}\left(L_4^\mathrm{r}{}^N-2L_6^\mathrm{r}{}^N\right)\right]\fs\\  
    \end{split}
\end{equation}
These results have already been obtained for $N=4$ (and with remarks on how to
proceed for general $N$) in \osn{2004} by P.~Hernandez and
M.~Laine\cite{Hernandez:2004ik}.

The matching of the $L_i^\mathrm{r}{}^{N-1}$ is
obtained by comparing the coefficients of the pertinent basis elements of
$\L_4^{N-1}$, leading to the leading order matching relations
\begin{equation}
  \label{eq:li}
  \begin{split}
    L_0^\mathrm{r}{}^{N-1} &= L_0^\mathrm{r}{}^N - \tfrac{1}{48}\nu_K\co\qquad
    L_1^\mathrm{r}{}^{N-1}  = L_1^\mathrm{r}{}^N\co\qquad
    L_2^\mathrm{r}{}^{N-1}  = L_2^\mathrm{r}{}^N\co\\
    L_3^\mathrm{r}{}^{N-1} &= L_3^\mathrm{r}{}^N - \tfrac{1}{24}\nu_K\co\qquad
    L_4^\mathrm{r}{}^{N-1}  = L_4^\mathrm{r}{}^N\co\qquad
    L_5^\mathrm{r}{}^{N-1}  = L_5^\mathrm{r}{}^N - \tfrac{1}{8}\nu_K\co\\
    L_6^\mathrm{r}{}^{N-1} &= L_6^\mathrm{r}{}^N + \tfrac{2N-1}{8N^2(N-1)^2}\nu_\eta-\tfrac{1}{128N(N-1)^2\pi^2}\co\\
    L_7^\mathrm{r}{}^{N-1} &= -\frac{F_N^2}{32(N-1)^2B_Nm} \\
&\quad+
    \tfrac{1}{(N-1)^2}\left(-L_4^\mathrm{r}{}^N+L_6^\mathrm{r}{}^N+N^2L_7^\mathrm{r}{}^N+L_8^\mathrm{r}{}^N\right)\\
&\quad-\tfrac{N-3}{16(N-1)^2}\nu_K-\tfrac{1}{8N^2}{\nu_\eta}+\tfrac{1}{512\pi^2}\left[\tfrac{2}{N^2}+\tfrac{N-3}{(N-1)^2}\right]\co\\
    L_8^\mathrm{r}{}^{N-1} &= L_8^\mathrm{r}{}^N - \tfrac{1}{4N(N-1)}\nu_\eta-\tfrac{1}{16}\nu_K+\tfrac{1}{128N(N-1)\pi^2}\co\\
    L_9^\mathrm{r}{}^{N-1} &= L_9^\mathrm{r}{}^N - \tfrac{1}{12}\nu_K\co\qquad
    L_{10}^\mathrm{r}{}^{N-1} = L_{10}^\mathrm{r}{}^N + \tfrac{1}{12}\nu_K\co\qquad
    L_{11}^\mathrm{r}{}^{N-1} = L_{11}^\mathrm{r}{}^N + \tfrac{1}{24}\nu_K\co\\
    L_{12}^\mathrm{r}{}^{N-1} &= L_{12}^\mathrm{r}{}^N - \tfrac{1}{2N(N-1)}\nu_\eta-\tfrac{1}{8}\nu_K+\tfrac{1}{64N(N-1)\pi^2}\co
  \end{split}
\end{equation}
where we used $\mu_P = \bar{M}_P^2/(32\pi^2)\ln(\bar{M}_P^2/\mu^2)$ and $\nu_P
= 1/(32\pi^2)\,[\ln(\bar{M}_P^2/\mu^2)+1]$ at an
arbitrary scale $\mu$ to represent the chiral logs with the tree--level masses of the particle $P$.

There are some checks available to the above result. The obvious one is to
check whether it reproduces the results for $N=3$ as obtained more
than a quarter of a
century ago by Gasser and Leutwyler\cite{glnpb}. Two obstacles have to be
overcome when performing this check. For one, the basis (\ref{eq:L4ops}) is
not minimal for $N=2$ or $N=3$. Two, the standard minimal basis for
\chpt$_2$ is not simply a reduced set of (\ref{eq:L4ops}), but is only
derived from it by the use of the equation of motion and a linear
combination of the other elements. Hence this check offers the opportunity to
show how to convert a result written in a nonminimal basis into a minimal one
which is not just a simple reduction of the former. This is done in Appendix~\ref{app:validity}.

Another check is to see whether the $\mu$--dependence on both sides of the
equations is the same. For this we need to recall that
\begin{equation}
  \label{eq:renorm}
  L^N_k = L^\mathrm{r}_k{}^N + \Gamma^N_k\lambda
\end{equation}
with a finite remainder $L^\mathrm{r}_k{}^N$ at $d=4$ and \cite{glnpb}
\begin{align}
  &\lambda =
  \frac{\mu^{d-4}}{(4\pi)^2}\left\{\frac{1}{d-4}-\tfrac{1}{2}[\ln(4\pi)+\Gamma'(1)+1]\right\}\co\nn\\
  \begin{split}
    &\Gamma^N_0=\tfrac{N}{48}\co\qquad
    \Gamma^N_1=\tfrac{1}{16}\co\qquad
    \Gamma^N_2=\tfrac{1}{8}\co\qquad
    \Gamma^N_3=\tfrac{N}{24}\co\qquad
    \Gamma^N_4=\tfrac{1}{8}\co\qquad\\
    &\Gamma^N_5=\tfrac{N}{8}\co\qquad
    \Gamma^N_6=\tfrac{N^2+2}{16N^2}\co\qquad
    \Gamma^N_7=0\co\qquad
    \Gamma^N_8=\tfrac{N^2-4}{16N}\co\qquad\\
    &\Gamma^N_9=\tfrac{N}{12}\co\qquad
    \Gamma^N_{10}=-\tfrac{N}{12}\co\qquad
    \Gamma^N_{11}=-\tfrac{N}{24}\co\qquad
    \Gamma^N_{12}=\tfrac{N^2-4}{8N}\fs
  \end{split}
\end{align}
While this check is simple, a large part of it (the determination of the
$\Gamma_k^N$) relies on the same technique as the calculation of the loop
contributions. Therefore, it is not as strong as one might guess first.


\section{Application}
\label{sec:application}

While an extension of the chiral perturbation theory concept beyond $N = 3$
for physical quark masses is rather far-fetched, it can be done for quarks on
the lattice. In our view, it is not enough to be able to simulate certain
effects at physical quark masses on a lattice, as many phenomena are either
hard to reach or bring their own specific problems along. One should in
addition try to anchor the simulations also in unphysical regions, where analytical
results are available. One such anchor can be provided with this paper: in
addition to the standard \chpt{} results, we can provide some information on
the charm quark mass dependence of the $\SU(3)$-\lecs{} in a region where all of
$m_u$, $m_d$, and $m_s$ are small compared to $m_c$ and the latter is itself
much smaller than $(4\pi F_0)^2/B_0$. In such a configuration,
the formalism of \chpt{} can be extended to four flavours and
Equations~(\ref{eq:li}) can be applied to extract the $m_c$--dependence. If
also the $m_s$--dependence is needed, one simply has to apply the
relations~(\ref{eq:li}) a second time, i.e. perform the double reduction
$\SU(4)\rightarrow\SU(3)\rightarrow\SU(2)$. Using this procedure, we can obtain
expressions for the ratios $F_3/F_4$, $B_3/B_4$ and $\Sigma_3/\Sigma_4$, where
$\Sigma_i = F_i^2B_i$. At the given order, occurrences of $F_4$ and $B_4$ on
the right--hand--side can be replaced with $F_3$ and $B_3$, the \lecs{} of
$\L_4^4$ can be translated to those of $\L_4^3$ by using the relations
(\ref{eq:li}) a second time. In the following, we adopt for the
$\SU(3)$--\lecs{} the conventional notation, i.e.  write $X_0$ for $X_3$ ($X
\in \{F;B;\Sigma\}$) and abbreviate $L_i^3{}^\mathrm{r}$ by
$L_i^\mathrm{r}$. Further we denote the charmed companions of $K$ and $\eta$
with $D$ and $\eta_c$, respectively. Their tree-level masses in the
$\SU(4)$--limit are $\bar{M}_D^2 = B_4m_c$ and $\bar{M}_{\eta_c}^2 =
\tfrac{3}{2}B_4m_c$. In that notation, we obtain to first order in $m_c$
\begin{equation}
  \label{eq:mc_lec}
  \begin{split}
  F_0/F_4 &= 1 -\frac{\mu_D}{F_0^2}+ 8\tfrac{B_0m_c}{F_0^2}L_4^\mathrm{r}\co\\
  B_0/B_4 &= 1 - \frac{\mu_{\eta_c}}{6F_0^2}- 16\tfrac{B_0m_c}{F_0^2}\left(L_4^\mathrm{r}-2L_6^\mathrm{r}+\tfrac{7}{576}\nu_{\eta_c}-\tfrac{1}{2304\pi^2}\right) \co\\
  \Sigma_0/\Sigma_4 &= 1 - 2\frac{\mu_D}{F_0^2} - \frac{\mu_{\eta_c}}{6F_0^2} + 32\tfrac{B_0m_c}{F_0^2}\left(L_6^\mathrm{r}-\tfrac{7}{1152}\nu_{\eta_c}+\tfrac{1}{4608\pi^2}\right)\fs
  \end{split}
\end{equation}

We plot these ratios up to $B_0m_c = 0.6\text{ GeV}^2$, where the expansion
parameter $B_0m_c/(4\pi F_0)^2$ is roughly 1/2. Note that this is still about
a factor four below the value obtained with physical charm quark
masses. Nevertheless, the error bars show that the predictive power of the
formul\ae has all but vanished at this point. We use parameters $F_0 =
87.2\text{ MeV}$, $L_4^\mathrm{r} = (0.0\pm0.5)\cdot 10^{-3}$, and
$L_6^\mathrm{r} = (0.0\pm0.3)\cdot 10^{-3}$ at the scale $\mu=M_\rho=770\text{
  MeV}$.

\begin{figure}[t!p]
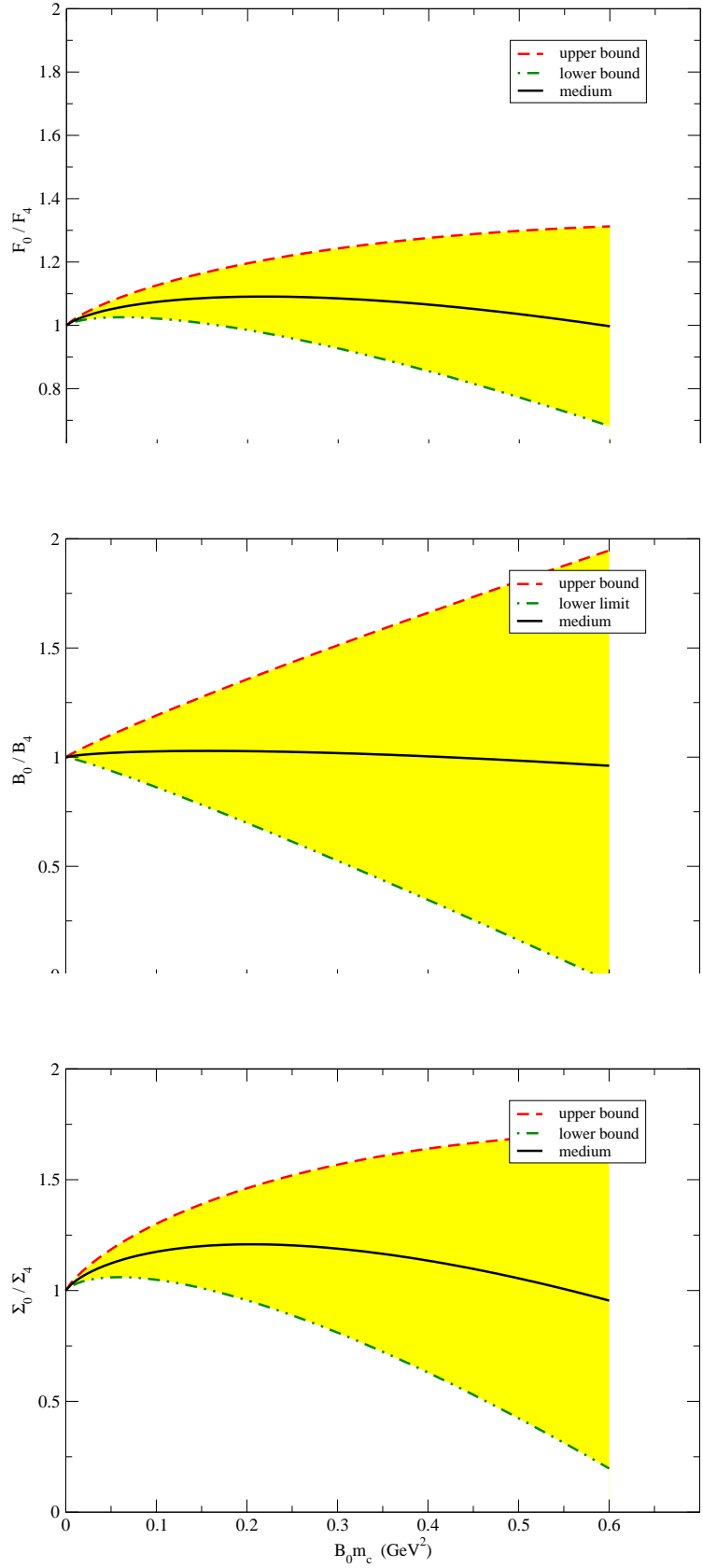

\centering
   \subfloat{\epsfig{file=F3F4_color.eps,height=15\baselineskip}}\\ 
   \subfloat{\epsfig{file=B3B4_color.eps,height=15\baselineskip}}\\
   \subfloat{\epsfig{file=S3S4_color.eps,height=15\baselineskip}}  
   \caption{Plots for the ratios $F_3/F_4$, $B_3/B_4$ and $\Sigma_3/\Sigma_4$.
}
   \label{fig:rat}
\end{figure}

In addition, the relations~(\ref{eq:li}) allow to determine the
$m_c$--dependence of known \chpt{}-results in pure $\SU(3)$--language at leading order from existing calculations. For example we
obtain for $m_u = m_d=\hat{m}$
\begin{align}
  \label{eq:mc_dep}
  \d_{m_c}\bra 0|\bar{q}q|0\ket &= B_0^2\left(2\nu_D+\tfrac{4}{9}\nu_{\eta_c}-32L_6^\mathrm{r}+\tfrac{82\hat{m}-7m_s}{m_c}\tfrac{1}{1152\pi^2}-\tfrac{1}{144\pi^2}\right)\mc\quad q\in\{u;d\}\co\\
  \d_{m_c}F_\pi &=
  \frac{B_0}{F_0}\left(-\nu_{D}+8L_4^\mathrm{r}-\tfrac{\hat{m}}{m_c}
    \tfrac{1}{32\pi^2}\right)\co\\
  \d_{m_c}M_\pi^2 &= \frac{2B_0^2\hat{m}}{F_0^2}\left[-\tfrac{4}{9}\nu_{\eta_c}-16(L_4^\mathrm{r}-2L_6^\mathrm{r})-\tfrac{10\hat{m}-7m_s}{m_c}\tfrac{1}{576\pi^2}+\tfrac{1}{144\pi^2}\right]\fs
\end{align}
To obtain these expressions, all one has to do is to substitute the \lecs{}
in the corresponding $\SU(3)$--expressions\cite{glnpb} with the pertinent
relations of~(\ref{eq:li}) to arrive at explicit formulae for the 
light quark mass dependence with \lecs{} that do not depend on these
masses. Afterwards, one can directly differentiate the formulae by $m_c$ and
apply the relations~(\ref{eq:li}) again to obtain expressions in the
$\SU(3)$--language. The dependence for other quantities can easily obtained by
this procedure form the existing literature.

 
\section{Summary}
\label{Sec:summary}

To summarise, we have determined the dependence on the $N^\text{th}$ quark mass
of the $N-1$--flavour \lecs{} $B_{N-1}$ and $F_{N-1}$ to next--to-leading and of the
\lecs{} $L_0^{N-1},\ldots,L_{12}^{N-1}$ to leading order. The calculation
relied on a matching between the local parts of the generating functionals of
\chpt${}_{N}$ and \chpt${}_{N-1}$. We hence showed that the same procedure used
for the determination of the strange quark mass dependence\cite{Gasser:2007sg,Gasser:2009hr,Haefeli:2007ey} of the two--flavour
\lecs{} can be generalised to the case with an arbitrary number of flavours.

These relations are useful to obtain constraints and further information on the
pertinent \lecs. We applied the relations to obtain the $m_c$--dependence of
the quark condensate in a limit where the mass of the charm quark is well
below half a GeV. This relation could be used in lattice calculations as an
additional analytic anchor in an unphysical regime.

\section*{Acknowledgements}
We thank H. Leutwyler for bringing our attention to this subject matter and
discussing it with us. M.A.I. appreciates the partial support of
the \caps{dfg} grant KO~1069/13-1 and
the  Russian Fund of Basic Research grant No.~10-02-00368-a.

\begin{appendix}
\section{Notation}
\label{app:notation}

In this appendix, we will settle the notation of \chpt$_N$, as this work is written in
Euclidean spacetime throughout. 

Chiral perturbation theory yields a consistent and systematic
framework to explore the effects of symmetries in low energy \qcd. The
starting point is the massless \qcd--Lagrangian $\L_\text{\qcd{}}^0$, enriched
with couplings to external (axial--) vector fields ($a_\mu$) $v_\mu$ and (pseudo--)
scalar sources ($p$) $s$.

Chiral perturbation theory is formulated with the aid of an effective Lagrangian
$\L_\text{\chpt{}} = \sum_n\L_{2n}$, where the degrees of freedom are the
emerging Goldstone bosons -- identified with the light mesons -- due to the
spontaneous symmetry breakdown inherent to $\L_\text{\qcd}^0$. The Lagrangian
densities $\L_{2n}$ are organised in a counting scheme that allows for an
expansion in the external momentum and the symmetry breaking terms. 

The mesons are
described in a Hermitian traceless field $\phi^N$, spanned by a basis 
of $N^2-1$ dimensions with elements $\{\lambda^N_a\}$, such that
$\phi^N=\phi^N_a\lambda^N_a$ (implicit summation over repeated indices is
assumed). As an explicit
representation, we choose one related to the fundamental representation,
but normalised as $\bra\lambda_a^N\lambda_b^N\ket = 2\delta_{ab}$.
There are $N-1$ purely diagonal elements 
\[\lambda_n^N = \sqrt{\frac{2}{n(n-1)}}\;\diag\big(1_{n-1},1-n,0_{N-n}\big)\mc\qquad
2\le n\le N\co\]
and $N(N-1)$ purely off--diagonal sparse Hermitian elements
\[\lambda_{nk}^N=i^{k\!\!\!\!\!\mod 2}\,e_{n\lceil k/2\rceil}+(-i)^{k\!\!\!\!\!\mod
  2}\,e_{\lceil k/2\rceil n}\mc\quad  n\in\{2;3;\ldots;N\}\co k\in\{1;2;\ldots;2(n-1)\}\fs\]
With $e_{kn}$ we denote the matrix whose only nonvanishing entry is a 1 in the
$k^\text{th}$ row and $n^\text{th}$ column. The particles described with
these matrices and their interactions considered below motivate to define
the following three subsets of the basis:
 The element
$\lambda_N^N$ will be denoted by $\lambda_\eta$, the $2(N-1)$ elements with
entries only in the $N^\text{th}$ row and column will be addressed by the set
$\lambda_K$ and all others by the set $\lambda_\pi$. This will simplify the
notation in the following, as the corresponding collection of particles are addressed with
the same labels $\eta$, $K$, and $\pi$. 
Any basis of traceless Hermitian $N\times N$-matrices with the above norm fulfils the
completeness relation
\begin{equation}
  \label{eq:complete}
  \sum_{a=1}^{N^2-1}(\lambda^N_a)_{kl}(\lambda^N_a)_{mn}=2\delta_{kn}\delta_{lm}-\tfrac{2}{N}\delta_{kl}\delta_{mn}\co
\end{equation}
which we will make use of later. 

We use a representation of the chiral symmetry where the operators $X$ of the Lagrangian transform under
chiral rotations $g=(g_L,g_R)\in \SU(N)_L\times\SU(N)_R$
as 
\begin{equation}
\label{eq:chirtrans}
X\rightarrow f(g,\phi^N)Xf(g,\phi^N)^{-1}\fs
\end{equation}
The compensator field $f$ is defined via the nonlinear representation of
the meson field $u^N(\phi^N)$ under the action of $g$ as
\[u^N(\phi^N)\rightarrow u^N(\phi^N{}')=g_Ru^N(\phi^N)f(g,\phi^N)^{-1} =
f(g,\phi^N)u^N(\phi^N)g_L{}^{-1}\fs\]
This representation is related to the customary $LR$--view via $U^N=(u^N)^2$, where
conventionally the explicit form of $U^N$ reads
$U^N=\exp{(i\phi^N/F_N)}$, with $F_N$ being one of the two
low-energy constants of the leading Lagrangian $\L_2$, carrying the dimension of
mass.

The elementary building blocks transforming as (\ref{eq:chirtrans}), used for
building the Lagrangians $\L_2^N$ and $\L_4^N$, are given
by 
\begin{equation}
\begin{split}
  u^N_\mu &= i\left[u^N{}^\dag(\d_\mu-ir^N_\mu)u^N-u^N(\d_\mu-il^N_\mu)u^N{}^\dag\right]\co\\
  \chi^N_\pm &= u^N{}^\dag\chi^N u^N{}^\dag\pm u^N\chi^N{}^\dag u^N\co\\
  f^N_{\pm\mu\nu}&=u^Nl^N_{\mu\nu}u^N{}^\dag\pm u^N{}^\dag r^N_{\mu\nu}u^N\co
  \end{split}
\end{equation}
where the following combinations of (axial--) vector sources
\[
  r^N_\mu = v^N_\mu+a^N_\mu\co\quad l^N_\mu = v^N_\mu-a^N_\mu\co\\
\]
and their field strengths
\begin{align*}
    r^N_{\mu\nu}&=\d_\mu r^N_\nu-\d_\nu r^N_\mu-i\comm{r^N_\mu}{r^N_\nu}\co\\
    l^N_{\mu\nu}&=\d_\mu l^N_\nu-\d_\nu l^N_\mu-i\comm{l^N_\mu}{l^N_\nu}\co
\end{align*}
as well as the (pseudo--) scalar source combination
\[
  \chi^N = 2B_N(s^N+ip^N)\co
\]
have been introduced.
Written in these building blocks, the leading order
Lagrangian reads
\begin{equation}
  \L_2^N = \frac{F_N^2}{4}\bra u^N\cdot u^N-\chi^N_+\ket\fs
\end{equation}
This Lagrangian implies the equations of motion
\begin{equation}
  \nabla^N_\mu u^N_\mu +\tfrac{i}{2}\tilde{\chi}^N_-=0\co
\end{equation}
where $\tilde{\chi}^N_-$ denotes the traceless part of $\chi^N_-$ and the
covariant derivative $\nabla^N_\mu$ is defined in terms of the chiral connection $\Gamma_\mu^N$ as
\begin{equation}
\begin{split}
\nabla^N_\mu\cdot{}&=\d_\mu\cdot{}+\,[\,\Gamma_\mu^N\co\,\cdot{}\;]\qquad\text{with}\\
\Gamma_\mu^N&=\tfrac{1}{2}\left[u^N{}^\dag(\d_\mu-ir^N_\mu)u^N+u^N(\d_\mu-il^N_\mu)u^N{}^\dag\right]\fs
\end{split}
\end{equation}

The generating functional $Z_N$ can be written in a series where the elements
are ordered by the number of loops involved in their determination. This
series is equivalent to reintroduce the old--fashioned $\hbar$ and expanding
$Z$ in powers of $\hbar$. The formal
derivation of this series relies on splitting the field $\phi$ into the part fulfilling the equations of motion
$\phi_\text{cl}$ and a quantum fluctuation $\xi$, parametrised as
\[
  U^N(\phi) = u^N_\text{cl}\,\exp \frac{i\xi}{F_N}\;u^N_\text{cl}\mc\qquad u^N_\text{cl} =
  u^N(\phi^N_\text{cl})\fs
\]
Counting $\xi$ as a quantity of order $\hbar^{1/2}$ and expanding the formal
representation of $Z_N$ as a path integral in powers of
$\hbar$ delivers the desired loop expansion,
\[Z^N=\bar{S}_2^N+\hbar\left(\bar{S}_4^N+\tfrac{1}{2}\ln\det D^N/D_0^N\right) + \ord{\hbar^2}\co\]
where $\bar{S}_n^N$ denotes the classical action belonging to $\L_n^N$ and the
differential operator $D^N$ is given in \chpt${}_N$ as
\begin{equation}
  \label{eq:diffop}
  D^N(x)=-d_x^2+\sigma^N(x)\mc\qquad
  d^x_\mu=\d^x_\mu+\hat{\Gamma}^N_\mu(x)\mc\qquad \left(D_0^N\right)_{ab}=-\delta_{ab}\left(\Delta+\bar{M}_a^2\right)\co
\end{equation}
with
\begin{equation}
\label{eq:operators}
\begin{split}
  \hat{\Gamma}^N_\mu{\,}_{ab}&=-\tfrac{1}{2}\bra\comm{\lambda^N_a}{\lambda^N_b}\Gamma^N_\mu\ket\co\\
    \sigma^N_{ab}&=\tfrac{1}{8}\bra\comm{\lambda^N_a}{u^N_\mu}\comm{\lambda^N_b}{u^N_\mu}\ket+\tfrac{1}{8}\bra\anticomm{\lambda^N_a}{\lambda^N_b}\chi^N_+\ket\fs
\end{split}
\end{equation}

The Lagrangian $\L_4^N$ is given as
\begin{equation}
  \L_4^N = \sum_{j=0}^{12}L^N_jX^N_j\co
\end{equation}
with \lecs{} $L^N_j$. The corresponding operators $X^N_j$ read\cite{glnpb}
\begin{equation}
  \label{eq:L4ops}
  \begin{split}
    X^N_0&=-\bra(u^N_\mu u^N_\nu)^2\ket\co&&\\
    X^N_1&=-\bra u^N\cdot u^N\ket^2,&\qquad 
        X^N_2&=-\bra u^N_\mu u^N_\nu\ket^2\co\\
        X^N_3&=-\bra(u^N\cdot u^N)^2\ket,&\qquad
        X^N_4&=\bra u^N\cdot u^N\ket\bra\chi^N_+\ket\co\\ 
        X^N_5&=\bra u^N\cdot u^N\,\chi^N_+\ket,&\qquad
        X^N_6&=-\bra\chi^N_+\ket^2\co\\ 
        X^N_7&=-\bra\chi^N_-\ket^2,&\qquad
        X^N_8&=-\tfrac{1}{2}\bra(\chi^N_+)^2+(\chi^N_-)^2\ket\co\\ 
        X^N_9&=\tfrac{i}{2}\bra f^N_{+\mu\nu}[\,u^N_\mu\,,\,u^N_\nu\,]\ket,&\qquad
        X^N_{10}&=-\tfrac{1}{4}\bra (f^N_+)^2-(f^N_-)^2\ket\co\\
    X^N_{11}&=-\tfrac{1}{2}\bra (f^N_+)^2+(f^N_-)^2\ket,&\qquad 
        X^N_{12}&=-\tfrac{1}{4}\bra(\chi^N_+)^2-(\chi^N_-)^2\ket\fs
    \end{split}
\end{equation}
We used the abbreviations $u^N\cdot u^N=u^N_\mu u^N_\mu$ and
$(f^N_\pm)^2=f^N_{\pm\mu\nu}f^N_{\pm\mu\nu}.$ 

\section{Tree-level}
\label{app:tree}

In this appendix we will point out the details of the tree--level contribution
calculation to the generating functional. As pointed out in
Section~\ref{sec:calculation}, the calculation involves the following steps:
\begin{itemize}
\item[-] show that $u^N$ is of the form (\ref{eq:tree_form})
\item[-] express the $\eta$ in terms of $\SU(N-1)$--fields
\item[-] show that the $\pi$ do not differ from those in $\SU(N-1)$ at the
  required order
\end{itemize}

To see the triviality of the classical $K$--fields, we observe that in our $N-1$--flavour limit the only fields
proportional to elements in $\lambda_K$ are the $K$--fields themselves. All
other fields and sources are proportional to a combination of elements of
$\lambda_\pi$, $\lambda_\eta$, or unity. Under the linear transformation
$\lambda_a^N\mapsto S\lambda_a^N S^{-1}$, with $S=\operatorname{diag}(1_{N-1},\,-1)$,
$\lambda_\pi$ and $\lambda_\eta$ are invariant, whereas the
elements of $\lambda_K$ pick
up a minus sign. However, $\L_2$ as a trace is invariant under the same
transformation, therefore also the equations of motion. Hence there can be
only vertices emitting an even number of $K$--lines. The invariance of the
Lagrangian under this transformation indicates a conservation law, known for $N=3$ as strangeness conservation.

But tree graphs containing $K$--particles cannot solely consist of vertices with an
even number of $K$--lines attached, as e.g.~the endpoints of the $K$--branches
of the tree have only one. Hence there are
no tree graphs with $K$--content in our $N-1$--flavour limit and the
solution of the equations of motion of the mesons can be written as a (commuting)
combination of $\pi$-- and $\eta$--fields:
\begin{equation}
  \label{eq:solfields}
  {u^N}=u^\pi e^{\tfrac{i}{2F_N}{\eta}\lambda_\eta}\fs
\end{equation}
Note that the field $u^\pi$ does not (necessarily) equal to $u^{N-1}$ of \chpt${}_{N-1}$, since it fulfils the equations of motion for the $\SU(N)$
version, which are different from the ones in the $N-1$--flavour case (see below).

Exploiting the simplification of the $N-1$--flavour limit in the representation of
the solution of the equations of motion, we may write the building blocks of the
Lagrangian $\L^N_2$ as
\begin{equation}
  \label{eq:app:build1}
  \begin{split}
    u^N_\mu&=u^\pi_\mu-\tfrac{1}{F_N}\lambda_\eta\d_\mu \eta\co\\
    \chi^N_\pm&=\frac{B_N}{B_{N-1}}\left(\chi^\pi_\pm\cos\alpha-i\chi_\mp^\pi\sin\alpha\right)+4B_Nm\,e_{NN}
    \begin{cases}
       \phantom{i}\cos (N-1)\alpha & \quad{\chi}_+^N\\
      i\sin (N-1)\alpha & \quad{\chi}_-^N
    \end{cases}\co\\
  \end{split}
\end{equation}
with $\alpha=\sqrt{2/[N(N-1)]}\,\eta/F_N$ and operators $X^\pi$ denote $X$ evaluated
with the fields $u^\pi$ and in the external fields only the $\SU(N-1)$--part
being different from zero, $B_N$ replaced by $B_{N-1}$. The only nonzero entry
of the matrix $e_{NN}$ consists of a 1 in the lower right corner. As will be shown in a moment, $\alpha$ is a quantity of order
$q^2,$ therefore we may expand the trigonometric functions for small $\alpha$ and
obtain a perturbation series up to a given order. Remarkably, the leading term
of the expansion of $\chi_+^N$ is not its $\SU(N-1)$ equivalent, but rather the
mass term $4B_Nme_{NN}$, which has a counting of $q^0$ and $\hbar^{-1}$. As we will see, this has
the effect that higher order terms of the $N$--flavour functional
contribute also to the leading term of the $N-1$--flavour theory.

Expressing the Lagrangian $\L_2^N$ in these terms, we can write down the
equation of motion for the $\eta$--particle as
\begin{equation}
\begin{split}
\left(\Delta-\bar{M}_\eta^2\right)\eta &=-\bar{M}_\eta^2\eta + F_NB_Nm\sqrt{\frac{2(N-1)}{N}}\sin(N-1)\alpha\\
&+\frac{F_N}{4}\frac{B_N}{B_{N-1}}\sqrt{\frac{2}{N(N-1)}}\left[\sin\alpha\bra\chi_+^l\ket+i\cos\alpha\bra\chi_-^l\ket\right]\co
\end{split}
\end{equation}
which can be solved for small $\alpha$ recursively.
Note that the sum of the first two terms on the right hand side is of the
order $\alpha^3$ in the expansion for small $\alpha$ when inserting the
tree-level mass (\ref{eq:tree_mass}). Therefore, the
inclusion of the $\eta$--mass term on the left hand side is natural. The
differential equation suggests a counting in which every occurrence of an
$\eta$--particle should count as $q^2$. We may now solve this equation
recursively for small momenta, respecting the counting, and find
\begin{equation}
  \label{eq:app:eta}
  \eta=-\frac{iF_N}{8B_{N-1}m}\sqrt{\frac{2N}{(N-1)^3}}\bra\chi^\pi_-\ket+O(q^4)\fs
\end{equation}

For the equations of motion of the $\pi$-particles we note that they are
different in the two theories in question,
\begin{align*}
  \SU(N)&:&\nabla_\mu^\pi u_\mu^\pi &= -\tfrac{i}{2}\tfrac{B_N}{B_{N-1}}\left(\tilde{\chi}_-^\pi\cos\alpha-i\tilde{\chi}_+^\pi\sin\alpha\right)\co\qquad\qquad\qquad\qquad\qquad\\
  \SU(N-1)&:&\nabla_\mu^{N-1} u_\mu^{N-1} &= -\tfrac{i}{2}\tilde{\chi}_-^{N-1}\fs
\end{align*}
 However, the difference is only a quantity of order
$q^4$, therefore we can assume for our application that the solutions are the
same, i.e.~fields with index $\cdot^\pi$ are equivalent to fields with index $\cdot^{N-1}$.

The counterterm contribution of $\L^N_4$ is evaluated at the solution of
the equation of motion also, therefore we can use the same technique to find
its representation in \chpt$_{N-1}$. This can be achieved by
using the following translation rule:
\begin{equation}
  \label{eq:f}
    {f}_{\pm\mu\nu}^N\rightarrow f^\pi_{\pm\mu\nu}\co
\end{equation}
to be understood to give the correct corresponding local term in the action
$\int\!\dmeas{x}{d}\,\L_4^N$.


\section{Loops at order $\hbar$}
\label{app:loop}

Having translated the tree-level part of the functional, we turn to the loop terms
of order $\hbar.$ Here we calculate again the contribution to the
$\SU(N)$--functional in our $N-1$--flavour limit to see how it is included in
pure \chpt$_{N-1}$. It
remains to calculate the local contributions of
\[Z_{1\,\mathrm{loop}}^N=\tfrac{1}{2}\ln\frac{\det D^N}{\det D^N_0}\fs\]

Following the work of Nyffeler and Schenk \cite{Nyffeler:1994ph} and
exploiting the
simplifications due to the $N-1$--flavour limit we are considering here
[$K$--particles do not mix with the others, consult~(\ref{eq:diffop})], the determinant can be 
written as
\begin{equation}
  \label{eq:app:det}
  \ln\det D^N = \ln\det D_{\pi}+\ln\det D_\eta+\ln\det
  D_K+\ln\det(1-D_\pi^{-1}D_{\pi\eta}D_\eta^{-1} D_{\eta\pi})\co
\end{equation}
where the index of $D^N$ denotes the subspace to
consider. 
Note that the inverses of the sub-blocks of $D^N$ do not correspond to the
propagators, since they do not include any mixing terms among massive and
massless particles. However, this makes them perfect candidates for a
treatment via heat--kernel techniques (for more detail consult
\cite{Nyffeler:1994ph} and the references therein). The mixing is here
explicitly present in the last term. We will now investigate each of these terms in turn. 

Following the argument that the corrections to the massless fields are
automatically generated by the generating functional of an effective
Lagrangian, we can neglect the part involving only $\pi$--particles, since it
will produce in the end the same terms as its $\SU(N-1)$ counterpart and will
hence be purely nonlocal.

The next two determinants describe tadpoles with insertions where only
particles of identical masses run in the loop: either $K$-- or
$\eta$--particles. Diagrams of this type can be represented (with internal
tree--level mass $z$) as
\begin{equation}
  \label{eq:tadpole}
  \ln\frac{\det D_z}{\det D_0} = -\sum_{n=1}^\infty
  \Gamma(n)F_n(z^2)\Tr\big(a_n\big)\co\qquad F_n(z^2)=\int\!\frac{\dmeas{\l}{d}}{(2\pi)^d}\,(z^2+\l^2)^{-n}\fs
\end{equation}
The first two Seeley-coefficients $a_1$ and $a_2$ are given as (with $\bar{\sigma} =
\sigma - z^2$)
\[
a_1 = -\bar\sigma\co \qquad 
a_2 = \tfrac{1}{2}\bar\sigma^2
-\tfrac{1}{6}\hat\nabla^2\bar\sigma
+\tfrac{1}{12}\hat{\Gamma}_{\mu\nu}\hat{\Gamma}_{\mu\nu}\co
\]
where the field strength $\hat{\Gamma}_{\mu\nu} =
[d_\mu\co\,d_\nu]$ and the differential $\hat{\nabla}_\mu\,\cdot = \d_\mu\cdot\,+\,[\,\hat{\Gamma}_\mu\co\,\cdot\,]$.
Note that each $a_n$ in the series is suppressed by an order $q^2$ to its
predecessor, therefore the method is tailored to our counting scheme.
The trace $\Tr$ is understood as a simultaneous ordinary trace and an integral
over position space. 

At order $q^4$, we get
\begin{align}
  \ln\frac{\det D_K}{\det D^0_K} &=
  F_1(\bar{M}_K^2)\Tr_K(\bar\sigma)-\tfrac{1}{12}F_2(\bar{M}_K^2)\Tr_K(6\bar\sigma^2+\hat\Gamma_{\mu\nu}\hat\Gamma_{\mu\nu})\co\\
  \ln\frac{\det D_\eta}{\det D^0_\eta} &= F_1(\bar{M}_\eta^2)\Tr_\eta(\bar\sigma) -
  \tfrac{1}{2}F_2(\bar{M}_\eta^2)\Tr_\eta(\bar\sigma^2)\fs
\end{align}
The field strength $\hat\Gamma_{\mu\nu}$ vanishes in the $\eta$--case. Also
note that the second term of $a_2$ is a total derivative and can therefore be
dropped. With
$\Tr_P$ we denote the partial flavour trace over the flavour subspace spanned
by the particle $P$ and an integral over position space.

In the above expressions, the only obstacle to overcome is the calculation of
the flavour traces. For the case of the $\eta$--particle, the trace is trivial as
here the Seeley--coefficients are one--dimensional. In the case of the
$K$--particle, however, matters are more involved. Formally, the problem is to
calculate the partial flavour trace over a product of matrices $X_{ab} =
X^1_{ac_1}\cdots X^{n+1}_{c_nb}$, where all indices are assumed to run only over the
subspace spanned by the $K$--particles. For the outer indices $a$ and $b$, the
solution is to calculate $\Tr_K = \Tr_N - \Tr_{N-1} - \Tr_\eta$. For the inner
indices, observe that in our $N-1$ flavour limit, any component
$X^k_{c_{k-1}c_k}$ with exactly one index in the $K$--space
vanishes. Therefore, the inner summation can be carried out over all
$\SU(N)$. Hence, all one needs to know is to calculate the trace over the
whole $\SU(N)$. This can be carried out the via the completeness relation
(\ref{eq:complete}). From this relation follows that
\begin{equation}
  \label{eq:completeness2}
  \begin{split}
    \sum_{a=1}^{N^2-1}\bra\lambda^N_aA\lambda^N_aB\ket&=-\tfrac{2}{N}\bra
    AB\ket+2\bra A\ket\bra B\ket\qquad\text{and}\\
    \sum_{a=1}^{N^2-1}\bra\lambda^N_aA\ket\bra\lambda^N_aB\ket&=2\bra
    AB\ket-\tfrac{2}{N}\bra A\ket\bra B\ket\fs
  \end{split}
\end{equation}
As the components of the Seeley--coefficients are given in (\ref{eq:operators}) in
exactly this basis, the traces can be determined mechanically.

The last term of Equation~(\ref{eq:det}) behaves somewhat differently, as it
contains two different propagators. We write
\begin{equation}
\label{eq:pe1l}
\ln \det (1-D_\pi^{-1}D_{\pi\eta}D_\eta^{-1}D_{\eta\pi}) = -\Tr(D_\pi^{-1}
D_{\pi\eta}D_\eta^{-1} D_{\eta\pi})+\text{non--local terms.}
\end{equation}
Note that already the first term of an expansion in the interaction of the determinant
yields all local contributions. Of course, the contained propagators
still have to be expanded via the heat--kernel techniques, but one has no
longer to bother which terms of the mixing contribution are local and which
aren't. At order $q^4$, the propagators amount just to their free variants and
the mixing vertex is $\bar\sigma_{\eta\pi}$.

The flavour traces and the translation of the vertices into terms of the
$N-1$--flavour theory can be performed by the same
methods as in the tree--level calculation.

All in all, the local contribution to $Z_\text{1loop}^N$ is of the form (\ref{eq:1loop:1}) 

For the occurring loop--integrals, we use the
$\overline{\text{MS}}$--scheme for their renormalisation. Multiplying
the $d$--dimensional loop--integrals with a factor $(\mu
c)^{-2\omega}$ yields the correct mass dimension when performing the
transition to four space--time dimensions. Here, we
introduced an arbitrary mass scale $\mu$ and a constant
$c$ that is conventionally chosen such that
\[\ln c = -\tfrac{1}{2}[\ln 4\pi +\Gamma'(1)+1]\fs\] 
The variation from four to $d$ space--time dimensions is contained in  
\[
\omega = d/2 - 2
\fs
\]
A Laurent--expansion of the resulting expressions around $\omega=0$ and
dropping the principal part yields the normalised results for the
loop--integrals. 

In $d$ dimensions, the needed loop--integrals read
\begin{equation}
\begin{split}
F_n^m(z)&=\int\!\frac{\dmeas{\l}{d}}{(2\pi)^d}\,\l^{-2m}(z+\l^2)^{m-n}\co
\qquad n>m\geq 0\co\qquad F_n(z)=F_n^0(z)\mc\\
&=z^{\omega+2-n}\,C(\omega)\,\frac{\Gamma(\omega+2-m)\Gamma(n-2-\omega)}
{\Gamma(\omega+2)\Gamma(n-m)}\co\qquad C(\omega)=(4\pi)^{-(2+\omega)}\fs
  \end{split}
\end{equation}


\section{Validity for $N = 3$}
\label{app:validity}
In this short appendix, we spell out the details of the check that our results
correspond to the ones already known for $N=3$ \cite{glnpb}.

Eliminating $X_0^3$ in the list of operators yields already the standard
minimal basis for $N=3$. Therefore, putting $L_0^\mathrm{r}{}^3=0$ in the results
(\ref{eq:li}) deals with the nonminimality of (\ref{eq:L4ops}) for $N=3$. Note
that this still leaves a loop contribution to $L_0^\mathrm{r}{}^2$, as a
kaon-loop with two insertions still produces a local contribution proportional
to $X_0^2$, as can be seen from (\ref{eq:1loop:2}).

The Cayley-Hamilton relation for two-dimensional matrices $A$ and $B$,
\[\{A\co\,B\} = A\bra B\ket + \bra A\ket B + \bra AB\ket - \bra A\ket\bra B\ket\co\]
can be used to yield a minimal basis for $N=2$ in the form of a reduced set of
(\ref{eq:L4ops}). In that manner we eliminate the elements $X_0^2$, $X_3^2$, and
$X_5^2$; i.e.~$X_l^2=\beta_{lk}X_k^2{}^\text{min}$. All remaining operators $X_k^2{}^\text{min}$ except for $X_4^2$ can be transformed
into the standard basis elements $K_j$ of \chpt$_2$\cite{glann} as linear combinations,
\begin{align*}
    K_1&=-\tfrac{1}{4}\bra u\cdot u\ket^2,&\qquad 
        K_2&=-\tfrac{1}{4}\bra u_\mu u_\nu\ket^2\co\\
        K_3&=-\tfrac{1}{16}\bra\chi_+\ket^2,&\qquad
        K_4&=\tfrac{i}{4}\bra u_\mu\chi_{-\mu}\ket\co\\ 
        K_5&=\tfrac{1}{2}\bra f_-^2\ket,&\qquad
        K_6&=-\tfrac{i}{4}\bra f_{+\mu\nu}[\,u_\mu\,,\,u_\nu\,]\ket\co\\ 
        K_7&=\tfrac{1}{16}\bra\chi_-\ket^2,&\qquad
        X_8&=-\tfrac{1}{8}(\det\chi_++\det\chi_-)\co\\ 
        K_9&=\bra f_+^2+f_-^2\ket\co&\qquad
        K_{10}&=-\tfrac{1}{16}\bra\chi_+^2-\chi_-^2\ket\fs
\end{align*} 
For simplicity, we neglected the superscript $\cdot^2$ to indicate that here
$N=2$ and introduced the notation
$\chi_{\pm\mu}=\nabla_\mu\chi_\pm-\tfrac{i}{2}\{\chi_\mp,u_\mu\}$. In this
notation, $\L_4=\sum_{j=1}^{10}l_jK_j$, where the commonly used $h_i$
translate to $h_1 - h_3 = l_8$, $h_2 = l_9$, and $h_3 = l_{10}$.

The element $X_4^2$ can be written as a combination of $K_4$ and other $K_j$
by
the use of the equations of motion. As the operators of the basis enter the
generating functional only at the point where the contained fields satisfy the
equations of motion, this is not an issue. One obtains in this manner a list
of transformation equations of the form $X_k^2{}^\text{min}=\alpha_{kj}K_j$. The
contribution to a specific \caps{lec}
$l_j^\mathrm{r}$ is then given by 
$l_j^\mathrm{r}=\beta_{lk}\alpha_{kj}L_l^\mathrm{r}{}^2$,
where for $L_l^\mathrm{r}{}^2$ the solutions of (\ref{eq:li}) are inserted.


\end{appendix}



\begin{thebibliography}{10}
\def\enquote#1{``#1''}
\expandafter\ifx\csname url\endcsname\relax
  \def\url#1{{\tt #1}}\fi
\expandafter\ifx\csname urlprefix\endcsname\relax\def\urlprefix{URL }\fi
\expandafter\ifx\csname eprint\endcsname\relax\def\eprint#1{\url{#1}}\fi

\bibitem{Weinberg78}
S.~Weinberg, \enquote{Phenomenological {L}agrangians}, {\em Physica\/} {\bf
  A96} (\oldstylenums{1979}), \oldstylenums{327}.

\bibitem{glann}
J.~Gasser and H.~Leutwyler, \enquote{Chiral perturbation theory to one loop},
  {\em Ann. Phys.\/} {\bf 158} (\oldstylenums{1984}), \oldstylenums{142}.

\bibitem{glnpb}
J.~Gasser and H.~Leutwyler, \enquote{Chiral perturbation theory: Expansions in
  the mass of the strange quark}, {\em Nucl. Phys.\/} {\bf B250}
  (\oldstylenums{1985}), \oldstylenums{465}.

\bibitem{Moussallam:2000zf}
B.~Moussallam, \enquote{{Flavor stability of the chiral vacuum and scalar meson
  dynamics}}, {\em JHEP\/} {\bf 08} (\oldstylenums{2000}), \oldstylenums{005},
  \eprint{hep-ph/0005245}.

\bibitem{Kaiser:2006uv}
R.~Kaiser and J.~Schweizer, \enquote{The expansion by regions in {$\pi K$}
  scattering}, {\em JHEP\/} {\bf 06} (\oldstylenums{2006}), \oldstylenums{9},
  \eprint{hep-ph/0603153}.

\bibitem{Gasser:2007sg}
J.~Gasser, C.~Haefeli, M.~A. Ivanov, et~al., \enquote{{Integrating out strange
  quarks in ChPT}}, {\em Phys. Lett.\/} {\bf B652} (\oldstylenums{2007}),
  \oldstylenums{21}, \eprint{0706.0955}.

\bibitem{Gasser:2009hr}
J.~Gasser, C.~Haefeli, M.~A. Ivanov, et~al., \enquote{{Integrating out strange
  quarks in ChPT: terms at order $p^6$}}, {\em Phys. Lett.\/} {\bf B675}
  (\oldstylenums{2009}), \oldstylenums{49}, \eprint{0903.0801}.

\bibitem{Gasser:2001un}
J.~Gasser, V.~E. Lyubovitskij, A.~Rusetsky, et~al., \enquote{{Decays of the
  $\pi_+\,\pi_-$ atom}}, {\em Phys. Rev.\/} {\bf D64} (\oldstylenums{2001}),
  \oldstylenums{16008}, \eprint{hep-ph/0103157}.

\bibitem{Jallouli:1997ux}
H.~Jallouli and H.~Sazdjian, \enquote{{Relativistic effects in the pionium
  lifetime}}, {\em Phys. Rev.\/} {\bf D58} (\oldstylenums{1998}),
  \oldstylenums{014011}, \eprint{hep-ph/9706450}.


\bibitem{Haefeli:2007ey}
C.~Haefeli, M.~A. Ivanov, and M.~Schmid, \enquote{{Electromagnetic low-energy
  constants in ChPT}}, {\em Eur. Phys. J.\/} {\bf C53} (\oldstylenums{2008}),
  \oldstylenums{549}, \eprint{0710.5432}.

\bibitem{Kampf:2009tk}
K.~Kampf and B.~Moussallam, \enquote{{Chiral expansions of the $\pi_0$ lifetime}},
  \eprint{0901.4688}.

\bibitem{tinu}
M.~Schmid, {\em Strangeless $\chi${\sc PT\rm} at large $m_s$\/}, Ph.D. thesis,
  University of Bern (\oldstylenums{2007}).

\bibitem{renorm}
J.~Bijnens, G.~Colangelo, and~G.~Ecker, \enquote{Renormalization of Chiral
  Perturbation Theory to Order $p^6$}, {\em Annals Phys.} {\bf 280}
{\oldstylenums{2000}}, \oldstylenums{100}, \eprint{hep-ph/9907333}

\bibitem{DescotesGenon:1999uh}
S.~Descotes-Genon, L.~Girlanda, and J.~Stern, \enquote{{Paramagnetic effect of
  light quark loops on chiral symmetry breaking}}, {\em JHEP\/} {\bf 01}
  (\oldstylenums{2000}), \oldstylenums{041}, \eprint{hep-ph/9910537}.

\bibitem{DescotesGenon:2002yv}
S.~Descotes-Genon, L.~Girlanda, and J.~Stern, \enquote{{Chiral order and
  fluctuations in multi-flavour QCD}}, {\em Eur. Phys. J.\/} {\bf C27}
  (\oldstylenums{2003}), \oldstylenums{115}, \eprint{hep-ph/0207337}.

\bibitem{DescotesGenon:2003cg}
S.~Descotes-Genon, N.~H. Fuchs, L.~Girlanda, et~al., \enquote{{Resumming QCD
  vacuum fluctuations in three-flavour chiral perturbation theory}}, {\em Eur.
  Phys. J.\/} {\bf C34} (\oldstylenums{2004}), \oldstylenums{201},
  \eprint{hep-ph/0311120}.

\bibitem{DescotesGenon:2007bs}
S.~Descotes-Genon, \enquote{{$\pi$-$\pi$ and $\pi$-$K$ scatterings in
    three-flavour 
  resummed chiral perturbation theory}}, {\em J. Phys. Conf. Ser.\/} {\bf 110}
  (\oldstylenums{2008}), \oldstylenums{052012}, \eprint{0710.1696}.

\bibitem{Brown}
D.~G. Boulware and L.~S. Brown, \enquote{Tree graphs and classical fields},
  {\em Phys. Rev.\/} {\bf 172} 5 (\oldstylenums{1968}), \oldstylenums{1628}.

\bibitem{Nyffeler:1994ph}
A.~Nyffeler and A.~Schenk, \enquote{Effective field theory of the linear
  {$\mathrm{O}(N)$} sigma model}, {\em Annals Phys.\/} {\bf 241}
  (\oldstylenums{1995}), \oldstylenums{301}, \eprint{hep-ph/9409436}.

\bibitem{Hernandez:2004ik}
P.~Hernandez and M.~Laine, \enquote{Charm mass dependence of the weak
  Hamiltonian in chiral perturbation theory}, {\em JHEP} {\bf 0409 }
(\osn{2004}),  \osn{018}, \eprint{hep-ph/0407086}.


\end{thebibliography}
\end{document}